# Phonon-induced anomalous gauge potential for photonic isolation in frequency space

Jianfan Yang[1], Luqi Yuan[2*], Tian Qin[1], Fangxing Zhang[1], Yao Chen[1], Xiaoshun Jiang[3], Xianfeng Chen[2], Shanhui Fan[4*] and Wenjie Wan[1, 2*]

[1]The State Key Laboratory of Advanced Optical Communication Systems and Networks,
University of Michigan-Shanghai Jiao Tong University Joint Institute,
Shanghai Jiao Tong University, Shanghai 200240, China
[2]MOE Key Laboratory for Laser Plasmas and Collaborative Innovation Center of IFSA,
School of Physics and Astronomy, Shanghai Jiao Tong University, Shanghai 200240, China
[3]National Laboratory of Solid State Microstructures,
College of Engineering and Applied Sciences, Nanjing University, Nanjing 210093, China
[4]Department of Electrical Engineering, Stanford University, Stanford, California 94305, USA
*Corresponding authors: Wenjie Wan wenjie.wan@sjtu.edu.cn,
Luqi Yuan yuanluqi@sjtu.edu.cn & Shanhui Fan shanhui@stanford.edu

**Photonic gauge potentials are crucial for manipulating charge-neutral photons like their counterpart electrons in the electromagnetic field, allowing analogous Aharonov-Bohm effect in photonics and paving the way for critical applications like photonic isolation. Normally, a gauge potential exhibits phase inversion along two opposite propagation paths. Here we experimentally demonstrate phonon-induced anomalous gauge potentials with non-inverted gauge phases in a spatial-frequency space, where quasi-phase-matched nonlinear Brillouin scatterings enable such unique direction-dependent gauge phases. Based on this scheme, we construct photonic isolators in the frequency domain permitting nonreciprocal propagation of light along the frequency axis, where coherent phase control in the photonic isolator allows switching completely the directionality through an Aharonov-Bohm interferometer. Moreover, similar coherent controlled unidirectional frequency conversions are also illustrated. These results may offer a unique platform for a compact, integrated solution to implement synthetic-dimension devices for on-chip optical signal processing.**

A gauge potential, which in nature couples only to charged particles, manifests in the intriguing the Aharonov-Bohm effects even in the absence of an electromagnetic field [1]. Recently, artificial effective gauge fields have been created for neutral particles like photons in real space [2-8] and also synthetic space [9-14], opening up an emerging field to mold the flow of photons just like their counterpart, electrons. The development of photonic gauge potentials enables significant applications involving unidirectional spatial propagation of light [7], nonreciprocal phase shift in photonic Aharonov-Bohm (AB) effect [2,3,15], robust bulk transport in photonic topological insulator [14], and chiral currents in synthetic dimensions [13]. A key motivation for demonstrating photonic gauge potential is to create effective magnetic fields for photons [10,11,16]. Such an effective magnetic field enables one to break reciprocity for photons and to achieve non-magnetic photonic isolation [10], which is crucial for on-chip photonic integration.

Previously, the pioneering works of photonic gauge potentials for photonic AB effect [3,15] utilized two standing-wave modulators with different modulation phases. A modulator can convert light between two photonic states with different frequencies. It was noted in Ref. [3] that a standing-wave modulator can provide a (gauge-dependent) nonreciprocal phase response. In the

Hamiltonian that describes the modulator, the phases associated with the up- and down-conversion processes are equal in magnitude to the modulation phase and opposite in sign [3]. Such a nonreciprocal phase response corresponds to a photonic gauge potential. Cascading two of such modulators with two different modulation phases remove the gauge dependency and allows for nonreciprocal transmission [2]. In another proposal, a travelling-wave spatiotemporal modulation can also induce nonreciprocity through interband/ intermode photonic transitions inside waveguide structures [17,18]. Such travelling-wave modulation, in particular, can be created with the use of acoustic phonons [19-22]. In these systems, a single travelling wave modulator directly creates a nonreciprocal intensity dynamics, since the frequency conversion process is phase-matched in the forward direction, and phase-mismatched in the backward direction [23-25].

In this paper, we experimentally demonstrate an anomalous gauge potential for photons. For this purpose, we consider an optomechanical system supporting both optical and acoustic resonance. Thus, an optical photon will also experience up- and down-conversion through the interaction with acoustic waves. However, in our system, the phases associated with the two time-reversed up- and down-conversion processes are not equal in magnitude due to a quasi-phase-matching condition in the nonlinear Stokes/ anti-Stokes scattering processes involving a travelling wave phonon. In contrast with Ref. [3], therefore, we refer such a scenario of having an unequal up- and down-conversion phases as having an anomalous gauge potential for photons. And we refer to such an optomechanical system as a gauge-field modulator. Based on this anomalous gauge potential, we demonstrate a photonic isolator in the frequency space, constructed as an interferometer with a parallel-connected pair of an electro-optic modulator and a gauge-field modulator (GM), which can achieve unidirectional frequency conversion. Moreover, we also demonstrate the direction of the frequency conversion can be coherently controlled by controlling these up-conversion and down-conversion phases. This compact direction-dependent GM enables nonreciprocal light transmission in the synthetic frequency domain, opening up new avenues for the synthetic dimension towards the practical applications in all-optical signal processing and on-chip photonic integration.

A general GM with an anomalous gauge potential is illustrated in Fig. 1a, where a photon undergoes a frequency conversion propagating through the GM with an internal and directional temporal dynamic modulation such as acoustic phonons. As a result, photons from the left and the right both experience frequency conversion upon propagating through the structure, but with different phases depending on the propagation direction. We emphasize that the effective gauge potentials generated in the GM are unique in the spatial-frequency space in our system. For example, in the forward Stokes process ($\omega_0 \rightarrow \omega_1 = \omega_0 - \Omega$), photon acquires the phase $\varphi_{Forward} = \int_0^1 A_F \cdot d\vec{l}_F$, where $A_F$ is the effective gauge potential in the spatial-frequency space for the forward Stokes process. Meanwhile, in the opposite backward direction, photon acquires the phase $\varphi_{Backward} = \int_1^0 A_B \cdot d\vec{l}_B$ for the backward anti-Stokes process ($\omega_0 \leftarrow \omega_1 = \omega_0 - \Omega$), where $A_B$ is the corresponding gauge potential. The anomalous direction-dependent gauge potential in the GM can be defined by $A_F \neq A_B$, such that $\varphi_{Backward} \neq -\varphi_{Forward}$. Here the photon conversion phases differ from the normal gauge potential constructed by traditional EMs where $\varphi_{Forward} = \int_0^1 A \cdot d\vec{l}_F = \int_1^0 A \cdot d\vec{l}_B = -\varphi_{Backward}$ with the same effective gauge potential $A$ [3,4].

The proposed GM can be utilized for a photonic isolator in the frequency space. In the spatial space, a photonic isolator allows light in one direction but blocks light in the opposite direction by breaking time reciprocity in an asymmetric scattering matrix [26]. Note that, here the two ports (waveguides or scattering channels) in or out of the isolator are confined in the same frequency state. In contrast, a photonic isolator in the frequency domain will permit light frequency conversion in one direction by forbidding it in the opposite one, however, between two ports at two different frequency states. Demonstrating such a photonic isolator in the frequency space is of importance especially in the context of recent interests in exploring synthetic frequency dimensions [27-33].

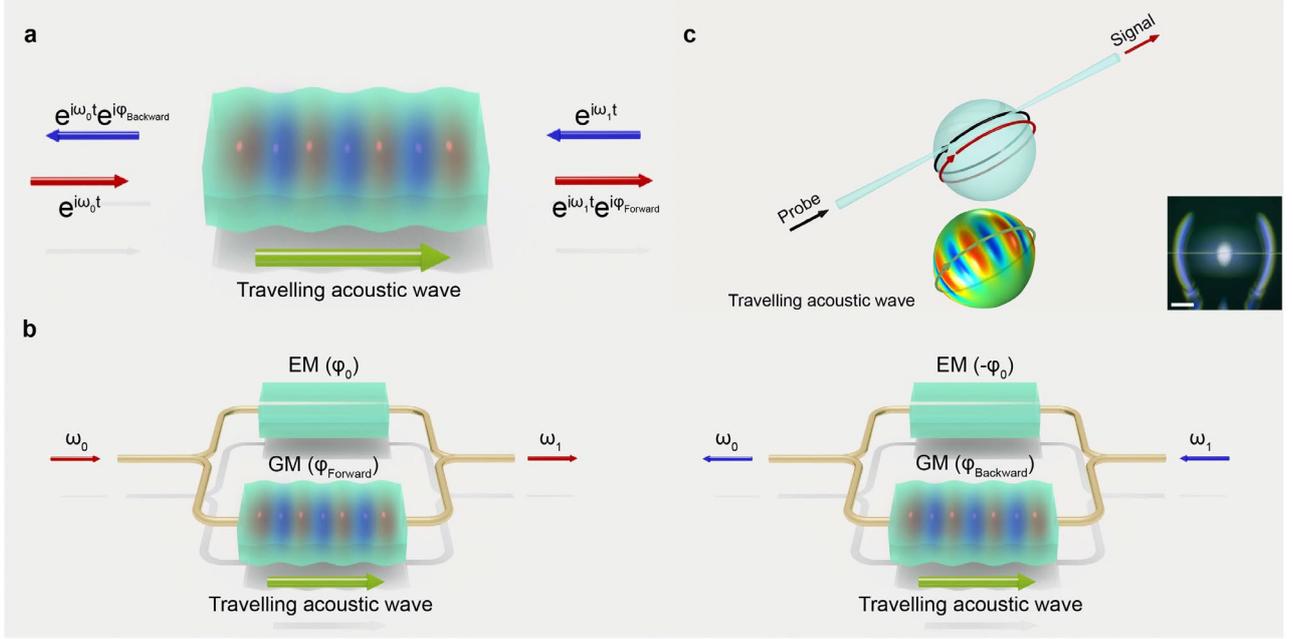

**Figure 1: Conceptual illustration of the gauge-field modulator with anomalous gauge potential and photonic isolator in frequency space.** (a) Working principle of the gauge-field modulator with nonreciprocal phase. A travelling acoustic wave propagates inside the device and directionally introduces a temporal dynamic modulation. Photon flows inject from two directions experience both a frequency shift and a phase shift which depends on whether the propagation direction of this photon flow is along or against the acoustic wave. (b) A photonic isolator in frequency space based on the interferometer scheme. A photon flow with a frequency of $\omega_0$/ $\omega_1$ propagating from the left/ right, up-converts/ down-converts to $\omega_1$/ $\omega_0$ and undergoes a phase shift $\varphi_0$/ $-\varphi_0$ in an electro-optical modulator (EM) and $\varphi_{Forward}$/ $\varphi_{Backward}$ in a gauge-field modulator (GM), and finally combines to interfere on the other side. (c) Schematic illustration of the gauge-field modulator based on a microsphere with forward probe light injection. Top: a probe light guided by a tapered fiber is evanescently coupled into a microsphere and experiencing the temporal dynamic modulation arising from the forward SBS phonon. This signal light is evanescently coupled back to the same tapered fiber. Bottom: a Rayleigh-type surface acoustic wave uni-directionally propagates in an acoustic WGM which is excited by the pump light at a different frequency. Inset: a photo of the microsphere coupling system. Scale bar: 20 μm.

To realize such a photonic isolator in the frequency space, we consider an interferometer with two separate frequency modulators/ converters including an EM and a GM as shown in Fig. 1b. Unlike a traditional EM during any frequency conversion, which exhibits a reciprocal phase in either spatial directions, i.e. $\varphi_{Forward} = \varphi_{Backward}$ and a normal gauge conversion phase in the frequency axis, i.e.

$\varphi_+ = -\varphi_-$ ($+/-$ for anti-Stokes and Stokes processes, respectively) [3], here the proposed GM (Fig. 1a) is path-depended in terms of directional unequal conversion phases, i.e. $\varphi_{Forward} \neq -\varphi_{Backward}$ for a reciprocal path loop in the spatial-frequency space (Fig. 1b) through a frequency conversion process $\omega_1 = \omega_0 \pm \Omega$. Meanwhile, the conversion phases inside the EM remain *opposite* to each other during the reciprocal processes, i.e. $\omega_0 \rightarrow \omega_1$ (Fig. 1b left) and $\omega_0 \leftarrow \omega_1$ (Fig. 1b right), which has been previously utilized for building normal gauge potential for photonic AB effect [3,15]. When such the GM and the EM are in a parallel connection, an effective gauge potential can be obtained in this spatial-frequency space such that photonic isolation in frequency conversion now is directional and phase-controllable through coherent interferences between the two arms. Note that, unlike the previous case of normal gauge potential in photonic AB effect [3,15], where two parallel-connected EMs each individually show the reciprocal conversion phases [3], i.e. $\varphi_0 = -\varphi_0$, during the anti-Stokes and Stokes processes in either forward or backward directions, ours exhibiting directional unequal conversion phases, i.e. $\varphi_{Forward} \neq -\varphi_{Backward}$, makes our gauge potential anomalous.

Experimentally, we can build such GMs based on the Brillouin scattering process associated with photon-phonon interactions as shown in Fig. 1c. Particularly, we consider a tapered fiber coupled microsphere resonator with $Q_{optical} > 10^8$ in the ambient environment [supplement]. Such a microsphere cavity can support both optical and acoustic whispering-gallery-type resonances, where a strong optical pump beam is launched into one optical WGM, simultaneously generating a Stokes photon with an extra acoustic travelling phonon through an electrostriction-induced stimulated Brillouin scattering (SBS) process [22,34]. Furthermore, the stimulated generated phonons effectively induce a travelling dynamic modulation necessary for a GM inside the cavity.

More physical insights on the GM can be understood by considering the internal phase-matching conditions in Fig. 2a: we consider that a probe photon at the frequency $\omega_0$ injects from the left undergoing a Brillouin scattering process and converts into $\omega_1 = \omega_0 \pm \Omega$, where $\pm \Omega$ depends on anti-Stokes/ Stokes processes involving phonon frequency $\Omega$. Normally, the phase-matching condition among two-photon and one phonon modes has to be fulfilled to ensure the conversion efficiency, for example, Stokes waves reverse their propagation directions with respect to their probes due to the large wave vector of high-frequency phonons (~ 11 GHz) during SBS inside optical fibers [35]. However, in our case, the wave vector of phonons $G$ is much smaller than that of photons (valid for low-frequency phonons ~100 MHz), i.e. $k \gg G$, a quasi-phase-matching regime can be obtained from the mismatching vector:

$$\Delta k = k_1 - k_0 \pm G \approx 0, \qquad (1)$$

where $k_i$ denotes the wave vector corresponding to the photons. Here a forward-type SBS [40] in cavity induces a forward-travelling phonon around 100 MHz, and its wave vector is much smaller than that of the photons centered at 1550 nm wavelength. As a result, wave vectors of the probe $k_0$ and of its anti-Stokes/ Stokes $k_1$ are at the same order of the magnitude, i.e. $k_0 \approx k_1$, due to this small frequency difference of phonons, similar to the case of acoustic optical modulator [15].

In this quasi-phase-matching regime, GMs can support both anti-Stokes and Stokes processes in both propagation directions due to the small phonon wave vector $G$ as shown in Figure 2. Take Stokes processes as an example: in the forward direction ($\omega_0 \rightarrow \omega_1 = \omega_0 - \Omega$), both the probe and the Stokes waves are aligned along the internal phonon's direction, and the momentum mismatching gives $\Delta k_{FSS} = k_0 - (k_1 + G)$. As a comparison, the backward Stokes ($\omega_0 - \Omega = \omega_1 \leftarrow \omega_0$) in the same GM gives $\Delta k_{BSS} = k_0 - (k_1 - G)$. One notices that $\Delta k_{FSS}, \Delta k_{BSS} \ll k_i$ according to the above argument, enabling the possibility for Stokes scattering generations in both directions, which is also verified experimentally later in Fig. 2b.

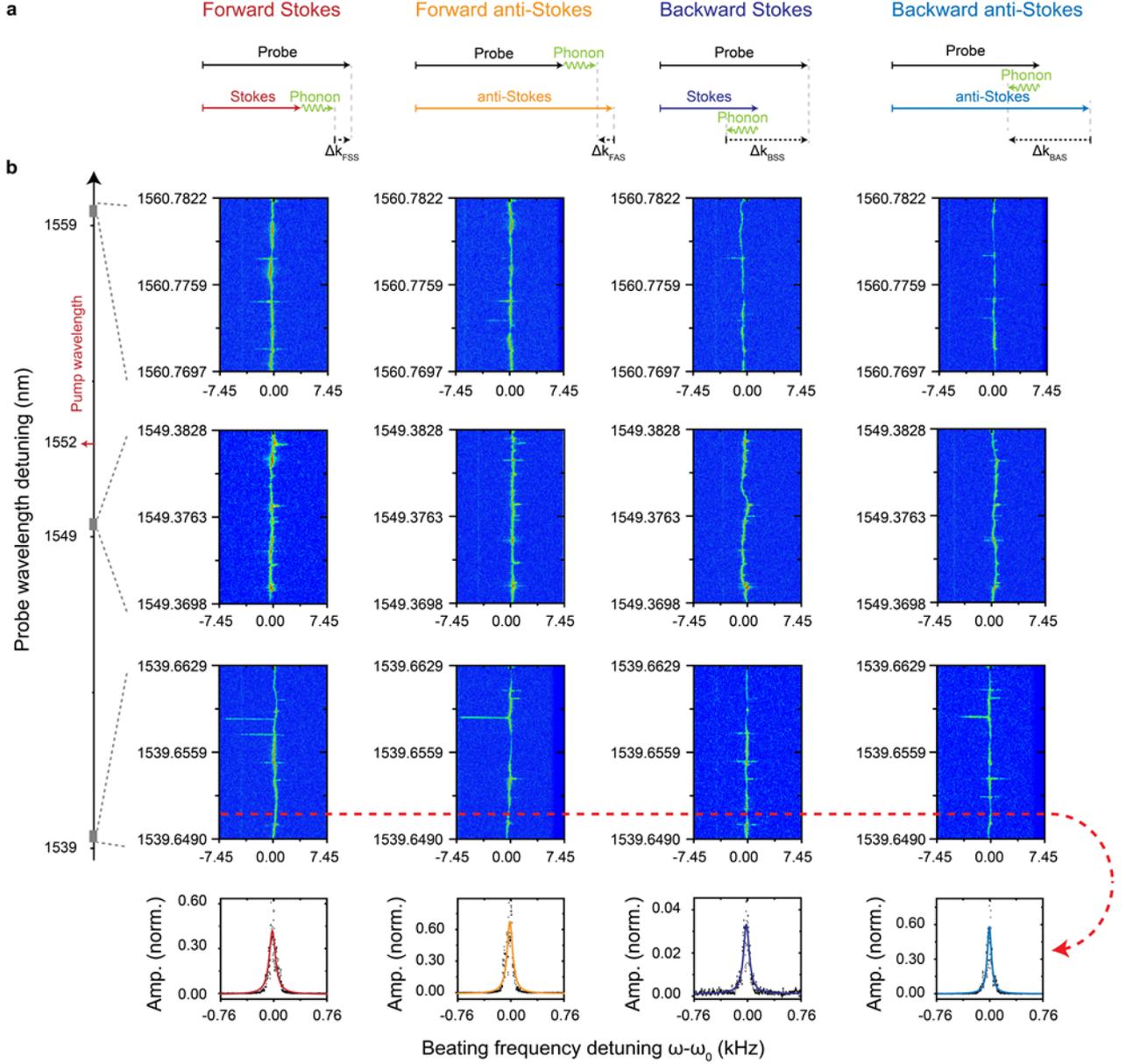

**Figure 2: Broadband quasi-phase-matched frequency conversion within the gauge-field modulator.** (a) Phase-matching diagrams of photon-phonon interaction in the GM. Forward Stokes/ anti-Stokes process and backward Stokes/ anti-Stokes process are illustrated in each diagram. The length of the arrow represents the value of the kinetic momentum of the photon/ phonon. The phase-mismatching term is represented by $\Delta k_{FSS}$, $\Delta k_{FAS}$, $\Delta k_{BSS}$, and $\Delta k_{BAS}$, respectively. (b) Experimental results for broadband temporal dynamic modulation. Probe light, at the blue-detuned (~1539 nm), near the pump wavelength (~1549 nm), and red-detuned (~1560 nm), shooting either in the forward direction or in the backward direction, experiences the forward SBS phonon (frequency at 507.220 MHz) and generates Stokes/ anti-Stokes signal. From left to right, each column describes the situation corresponding to forward Stokes

process, forward anti-Stokes process, backward Stokes process, and backward anti-Stokes process. The first three rows illustrate the trend of the generated beating signal versus probe light wavelength detuning in three wavelength domains. The last row shows the beating signal amplitude versus beating frequency detuning at the probe wavelength of 1539.6510 nm. The crossing dot is the experimental data, and the color line is the Lorentz fitting. Offset from (i) to (xvi): 507.220MHz + (0.00; 0.00; 0.00; 0.00; -1.86; -2.79; -0.93; -1.86; 0.00; -0.93; 0.00; -0.93; 0.06; -0.41; -0.58; -0.76) kHz. Color label: red for forward Stokes process; yellow for forward anti-Stokes process; blue for backward Stokes process; light blue for backward anti-Stokes process.

Importantly, this difference between these two mismatching vectors reveals a hidden feature of *nonreciprocal phase* during the frequency conversions: for the forward (left to right) Stokes process (Fig. 2a), the Stokes wave gains additional conversion phase as $e^{i\varphi_{FSS}}$, where $\varphi_{FSS} = \Delta k_{FSS} L = \int_0^1 A_F \cdot d\vec{l}_F$, L is the GM's path length, and $A$ is the effective gauge field [supplement]. Moreover, for the forward anti-Stokes process, the additional phase is $e^{i\varphi_{FAS}}$, where $\varphi_{FAS} = \Delta k_{FAS} L = \int_1^0 A_F \cdot d\vec{l}_F$. We notice that $\Delta k_{FSS} = -\Delta k_{FAS}$, which leads to $\varphi_{FSS} = -\varphi_{FAS}$. Therefore, one can safely define the effective gauge potential $A_F$ along the frequency axis of light for the forward propagation. Meanwhile, we can obtain the conversion phase from a backward Stokes process ($\omega_0 - \Omega = \omega_1 \leftarrow \omega_0$) as $\varphi_{BSS} = \Delta k_{BSS} L = \int_0^1 A_B \cdot d\vec{l}_B$. Furthermore, the backward anti-Stokes process ($\omega_0 \leftarrow \omega_1 = \omega_0 - \Omega$) associated with the photonic isolator mentioned above poses a conversion phase of $\varphi_{BAS} = \Delta k_{BAS} L = \int_1^0 A_B \cdot d\vec{l}_B = -\varphi_{BSS}$ due to $\Delta k_{BAS} = (k_1 - G) - k_0 = -\Delta k_{BSS}$, which indicates a crucial conclusion that the backward Stokes and anti-Stokes processes support the same effective gauge potential $A_B$ along the frequency axis of light for the backward propagation.

Moreover, there clearly exists a nonzero nonreciprocal phase difference between the forward and backward Stokes processes (to be verified experimentally in Fig. 3):

$$\Delta \varphi = (\Delta k_{BSS} - \Delta k_{FSS})L = 2GL \neq 0. \qquad (2)$$

Therefore, our GM support the path-dependent effective gauge potentials, i.e., $A_F \neq A_B$. In contrast, normal EMs pose the same phase for the Stokes process from either direction, which cannot provide the necessary nonreciprocal phase in the frequency space [supplement]. Similarly, such nonreciprocal phases also hold up for anti-Stokes processes in the GM. Note that, unlike previous cases of phase-matched scenarios where frequency conversions are only permitted in one phase-matched direction while inhibited in the other [19-21], here frequency conversions are allowed in both directions thanks for such *bidirectional quasi-phase-matching*, meanwhile, nonreciprocal phases are induced by directional phonon modulations in our GMs.

Experimentally, we exam anti-Stokes and Stokes frequency conversions for both propagation directions over a wide range (1539 nm-1561 nm) within the GM as shown in Fig. 2b. These conversions are widely continuous across all the frequency spectra we tested, verifying the nature of the quasi-phase-matching condition in Eq. (1) due to small mismatching vector $\Delta k$. Moreover, this condition can be also applied to both anti-Stokes/ Stokes processes, as well as forward and backward directions (Fig. 2b), making the frequency conversions possible to experimentally observe in all these cases (Fig. 2b). However, these conversions are companying with a low conversion efficiency estimated to be less than 0.01%, unlike other phase-matching cases [19]. Furthermore, we also observe some resonant spikes in all the cases, where either the optical resonances near the probe or the Stokes/anti-Stokes can enhance these conversions resulting in spike patterns on the beating spectra in Fig. 2b, which requires some further studies for resonant-enhanced conversion efficiency [supplement].

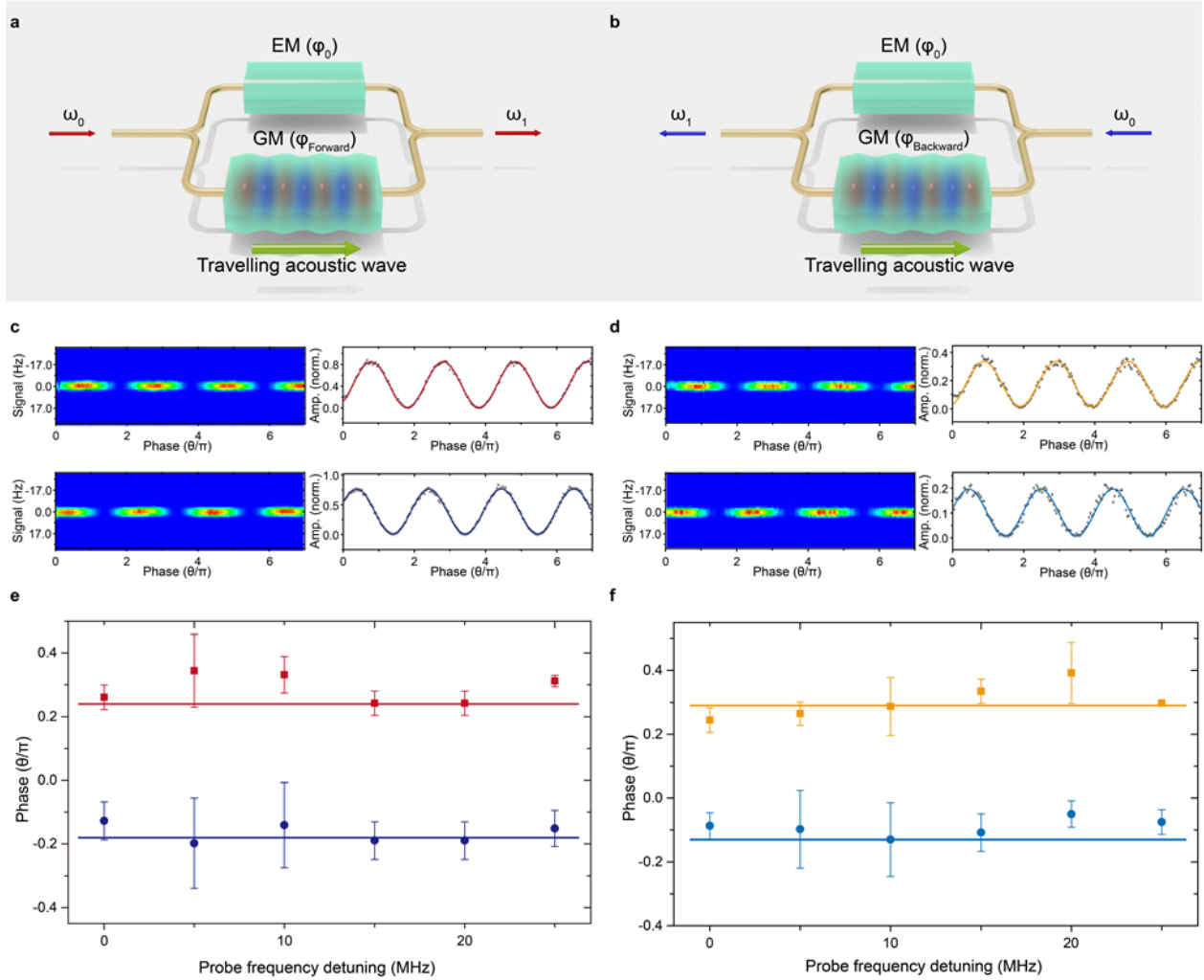

**Figure 3: Nonreciprocal conversion phase in a gauge potential loop.** Schematic of Mach-Zehnder interferometer for nonreciprocal phase verification in forward (a) and backward direction (b). Interference between Stokes signal (c) / anti-Stokes signal (d) and the EM modulated light in the forward direction (Top) and backward direction (Bottom). The first column is the experimental result for interference power evolution versus EM phase detuning. The second column records the peak power of the interference from the first column. The initial phase of the interference versus the probe frequency detuning in each case is recorded in (e) and (f). The gap of the linear fitting implies a constant nonreciprocal phase both existing and preserving the same value in the Stokes process and anti-Stokes process. Forward SBS phonon frequency (offset): 66.502 MHz. Color label: red for forward Stokes process; yellow for forward anti-Stokes process; blue for backward Stokes process; light blue for backward anti-Stokes process.

As mentioned above, the key ingredient in our GM to support gauge potential is the directional nonreciprocal phases. Based on an AB interferometer loop (Fig. 3a) interfering with an EM and a GM, we can effectively exam such nonreciprocal phases during frequency conversions for both directions. Both Stokes (Fig. 3c) and anti-Stokes (Fig. 3d) exhibit nonreciprocal phases between forward and backward propagations by comparing their interference patterns, which is done by fully scanning the phase spectrum through the external controlled EM. Furthermore, the relative differences between the forward and backward cases stay the same level for both the Stokes and anti-Stokes processes (Fig. 3e-f) across the probe's frequency detuning (~20 MHz). This could be understood from the above argument in Eq. (2), which explicitly indicates the origin of phonon induced nonreciprocal phase depending only on phonon's wave vector $G$. In the following, we will explore this nonreciprocal phase in GMs to construct effective gauge potential for unique applications in unidirectional frequency converters and photonic isolators for synthetic frequency space.

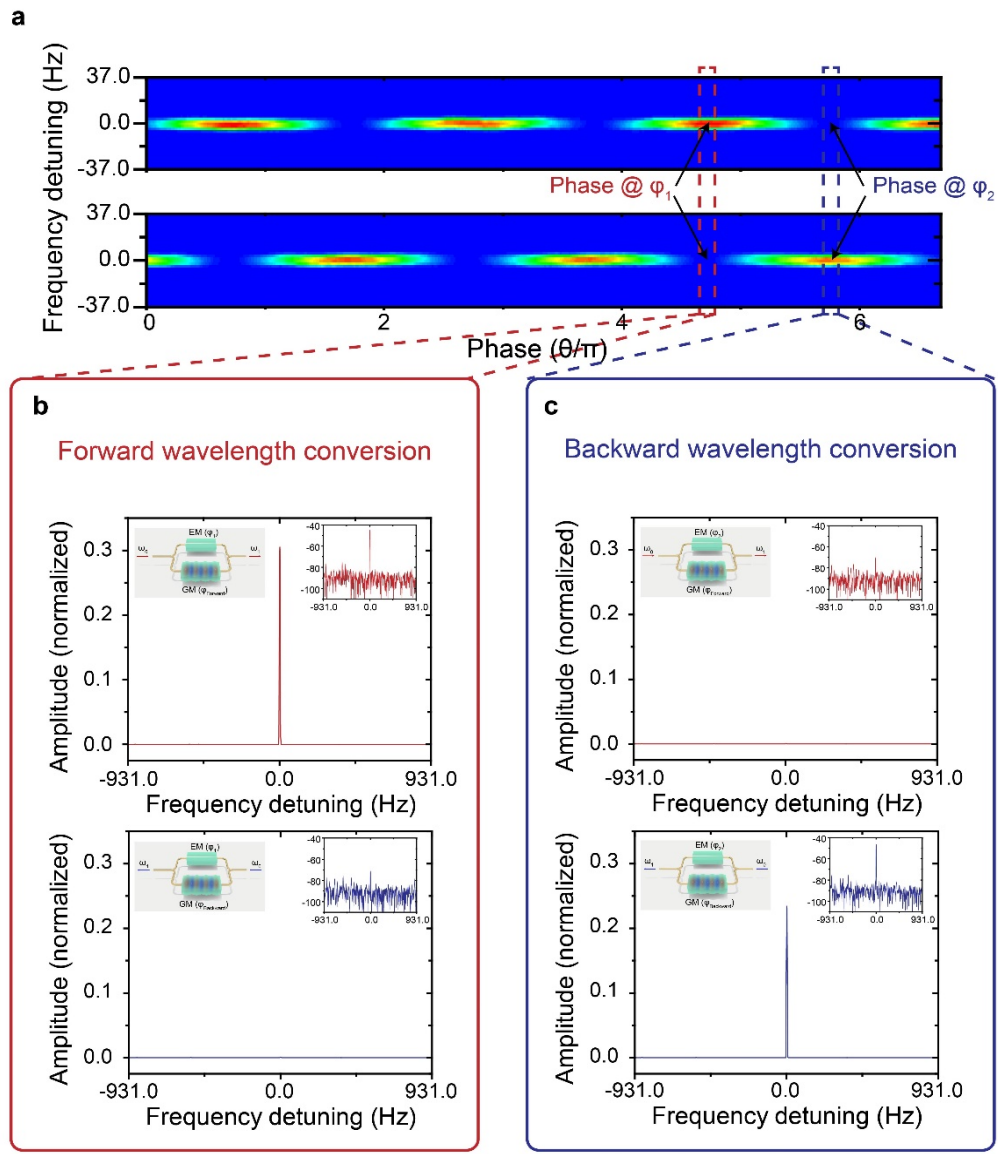

**Figure 4: Experimental demonstration of a coherent controlled unidirectional frequency converter.** (a) A typical pair of waterfall diagrams of interference between the Stokes signal and the EM modulated light in the forward direction (Top) and backward direction (Bottom). The initial phase difference is around 1.05π. Forward wavelength conversion at phase $\varphi_1$ with contrast ratio 26 dB (b) and backward wavelength conversion at phase $\varphi_2$ with contrast ratio 26 dB (c) are shown. Diagrams of the amplitude of the beating signal between the probe and the Stokes signal versus probe frequency detuning in linear scale are given both in the forward direction (Top) and the backward direction (Bottom). Data in log scale are provided in the inset. Forward SBS phonon frequency (offset): 141.740 MHz. Color label: red for forward Stokes process; blue for backward Stokes process.

Figure 4 shows an application of a unidirectional frequency converter based on the same configuration of Figure 3. For input photons at the frequency $\omega_0$, we first consider the case that the photon is injected at the left side, and the interference signal out of the parallel-connected EM and GM is transmitted at the right side. In this forward Stokes setup, the phase difference at the right side is $\Delta\phi_+ = \Delta k_{FSS}L + \varphi$, while, for the backward Stokes case, the phase difference at the left side is $\Delta\phi_- = \Delta k_{BSS}L + \varphi$. Here the reciprocal phase $\varphi$ is the EM's modulation phase, which can be externally controlled through the EM (Fig. 4a). Therefore, interference between the paths with EM and GM leads to a flexible way for coherent controlling frequency conversion through active phase control. One can purposely select a regime where the conversion in the forward direction is constructive while the conversion in the backward direction is destructive, resulting in unidirectional frequency conversion in the forward direction as shown in Fig.4b. More than 20 dB signal contrast of unidirectional conversions can be obtained in the experiment. Meanwhile, with an additional π phase added to the EM, such conversion can be totally reversed into the backward direction while inhibiting the forward conversion with a similar signal contrast (Fig. 4c).

Similarly, a photonic isolator in the frequency space can be demonstrated on the same interferometer platform in Figure 5. As proposed in Fig. 1, we can also observe a nonreciprocal frequency conversion between a forward Stokes ($\omega_0 \to \omega_1$) and its corresponding backward anti-Stokes ($\omega_0 \leftarrow \omega_1$) because of $\Delta k_{FSS} \neq \Delta k_{BAS}$ as discussed above. Moreover, similar to the unidirectional converter case, such a photonic isolator can be coherently controlled through the EM's phase, making its isolation direction switchable as shown in Fig. 5 b,c. The signal contrast is also more than 20 dB. The photonic isolators ensure the nonreciprocal light propagation in the spatial domain (forward and backward) as well as the frequency one ($\omega_0, \omega_1$), opening up a new door for all-optical signal processing in the synthetic dimension.

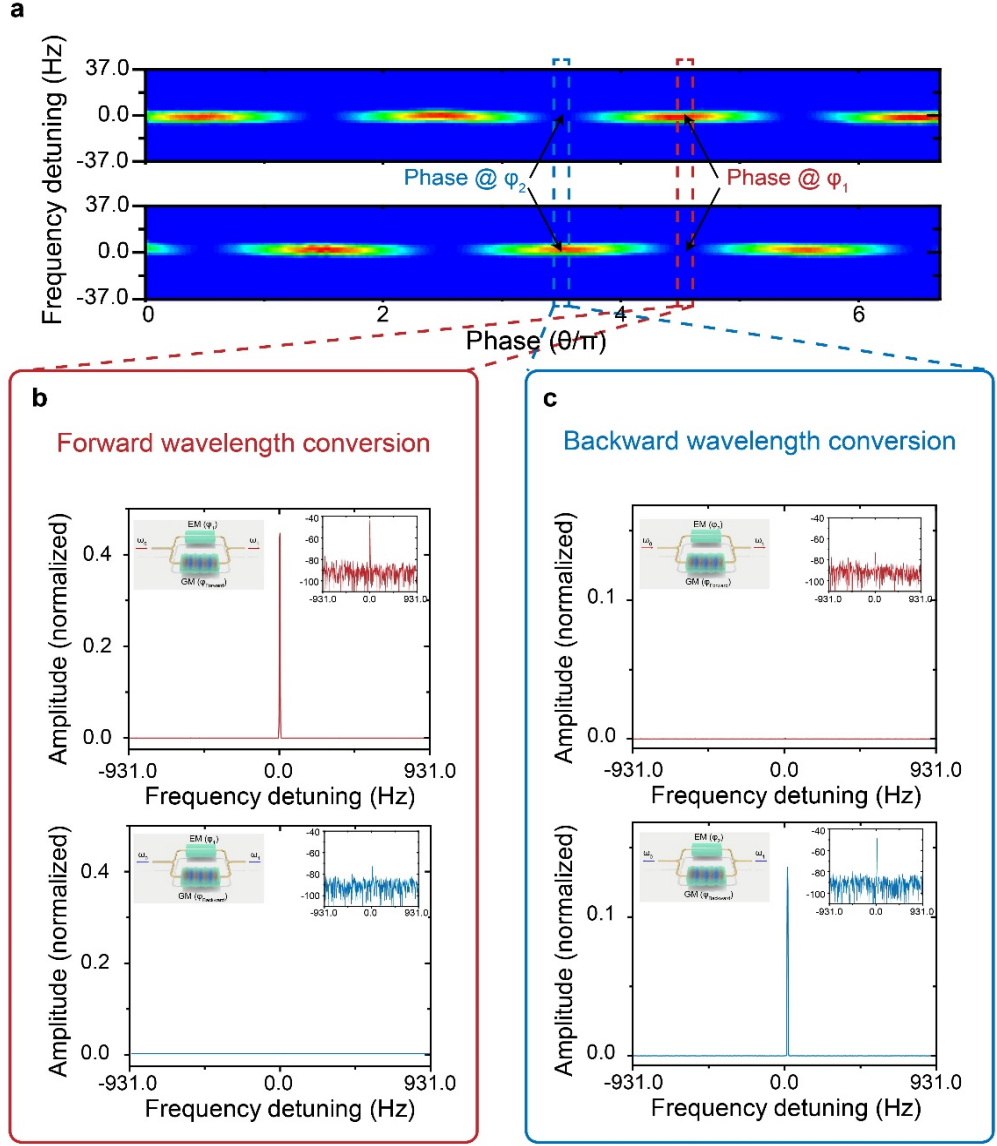

**Figure 5: Experimental demonstration of a coherent controlled photonic isolator in the frequency space.** (a) A typical pair of waterfall diagrams of interference between Stokes signal (Top)/ anti-Stokes signal (Bottom) and the EM modulated light in the forward direction (Top) and backward direction (Bottom). Forward wavelength conversion at phase $\varphi_1$ with contrast ratio 29 dB (b) and backward wavelength conversion at phase $\varphi_2$ with contrast ratio 24 dB (c) are shown. Diagrams of the amplitude of the beating signal between the probe and the Stokes signal versus probe frequency detuning in linear scale are given both in the forward direction (Top) and the backward direction (Bottom). Data in log scale are provided in the inset. Forward SBS phonon frequency (offset): 141.740 MHz. The probe frequency is 1539.4097 nm (for Stokes) and 1539.4102 nm (for anti-Stokes), respectively.

At last, we would like to discuss the potential applications of such phase-controlled unidirectional converters and photonic isolators. Like the prominent case of multi-wavelength signal processing in optical communications, i.e. dense wavelength division multiplexing (DWDM), on-chip multi-frequency signal processing promises a bright future for dramatically increasing signal bandwidths and processing capacity [36,37]. Here optical signals may need to be constantly converted back and forth in the frequency domain using nonlinear optics, which is a contrast to DWDM techniques where individual wavelengths serve as parallel and independent channels for transmitting signals simultaneously. Like the important on-chip element of photonic isolators for single-wavelength photonic integration, this unidirectional converter and the photonic isolator can play the same role in such multi-frequency on-chip signal processing preventing signals penetrating unwanted directions and frequency channels. However, it still poses a great challenge to break the reciprocal phases for frequency conversions, which is extremely crucial for recent exploration in synthetic dimensions [27], where physics can exhibit high-dimensional physical phenomena within low-dimensional photonic structures by using a different degree of freedom of photons, including optical frequency [9,11,28-30], and others[31-33]. All of them hold the potential for miniaturization towards on-chip applications [38-42], similar to dense-wavelength division multiplexing (DWDM) in fibers. More importantly, they can also be coherently controlled in directions offering more functionality such as signal routers or switches. Our proposed GM provides an alternative modulator for either the electro-optic modulation [42] and the four-wave-mixing process [29] which has been used to construct the synthetic dimension along the frequency axis of light. Unique direction-dependent modulation phases imprinted in GM give a realistic experimental platform to explore further researches in the spatial-frequency synthetic space, which is of significant emerging interest in the field of topological photonics.

**Acknowledgments:**

This work was supported by the National key research and development program (Grant No. 2016YFA0302500, 2017YFA0303700); National Science Foundation of China (Grant No. 11674228, No. 11974245, No. 11304201). Shanghai MEC Scientific Innovation Program (Grant No. E00075); S. F. acknowledge the support of U. S. National Science Foundation (Grant No. CBET-1641069)